\documentclass[useAMS,usenatbib,usegraphicx]{mn2e}
\usepackage{rotating}

\newcommand{\xte}{{\it RXTE}}
\newcommand{\chan}{{\it Chandra}}
\newcommand{\xmm}{{\it XMM-Newton}}

\title[Optical and X-ray variability of SXP46.6 and SXP6.85]{Optical and X-ray variability of two Small Magellanic Cloud X-ray binary pulsars - SXP46.6 and SXP6.85}
\author[K. E. McGowan et al.]{K. E. McGowan$^{1}$\thanks{E-mail:
kem@astro.soton.ac.uk}, M. J. Coe$^{1}$, M. P. E. Schurch$^{1}$, 
R. H. D. Corbet$^{2}$,
\newauthor
J. L. Galache$^{3}$, A. Udalski$^{4}$ \\
$^{1}$School of Physics and Astronomy, Southampton University, Highfield, 
Southampton, SO17 1BJ \\
$^{3}$University of Maryland, Baltimore County, Mail Code 662, NASA Goddard 
Space Flight Center, Greenbelt, MD 20771, USA \\
$^{3}$Harvard-Smithsonian Center for Astrophysics, Cambridge, MA 02138, USA \\
$^{4}$Warsaw University Observatory, Aleje Ujazdowskie 4, 00-478 Warsaw,
Poland}

\begin{document}

\date{}

\pagerange{\pageref{firstpage}--\pageref{lastpage}} \pubyear{2007}

\maketitle

\label{firstpage}

\begin{abstract}
We present long-term optical and {\it RXTE} data of two X-ray binary pulsars 
in the Small Magellanic Cloud, SXP46.6 and SXP6.85.  The optical light curves 
of both sources show substantial ($\sim0.5-0.8$ mag) changes over the time span
of the observations.  While the optical data for SXP6.85 do not reveal any
periodic behaviour, by detrending the optical measurements for SXP46.6 we
find an orbital period of $\sim 137$ days, consistent with results from the
X-ray data.  The detection of Type I X-ray outbursts from SXP46.6, combined 
with the fact that we also see optical outbursts at these times, implies that
SXP46.6 is a high orbital eccentricity system.  Using contemporaneous optical 
spectra of SXP46.6 we find that the equivalent width of the H$\alpha$ emission 
line changes over time indicating that the size of the circumstellar disc 
varies.  By studying the history of the colour variations for SXP6.85 we find 
that the source gets redder as it brightens which can also be attributed to 
changes in the circumstellar disc.  We do not find any correlation between the 
X-ray and optical data for SXP6.85.  The results for SXP6.85 suggest that it 
is a low eccentricity binary and that the optical modulations are due to the 
Be phenomenon.
\end{abstract}

\begin{keywords}
X-rays: binaries -- stars: emission-line, Be -- (galaxies:) Magellanic Clouds
\end{keywords}

\section{Introduction}

Observations of the Small Magellanic Cloud (SMC) at multi-wavelengths have
uncovered numerous high-mass X-ray binaries, the majority of which are 
Be/X-ray transients \citep[see e.g.][]{hab04,coe05,mcg07}.  These systems 
comprise of an OB star and a neutron star.  As the compact object orbits the 
companion star in a wide and eccentric orbit it passes close to the 
circumstellar disc of the Be star and X-ray outbursts are seen \citep{oka01}.  
In some cases optical outbursts coincident with the X-ray outbursts are also 
detected \citep[see e.g.][]{alc01,coe04}.

\begin{figure}
 \includegraphics[width=84mm]{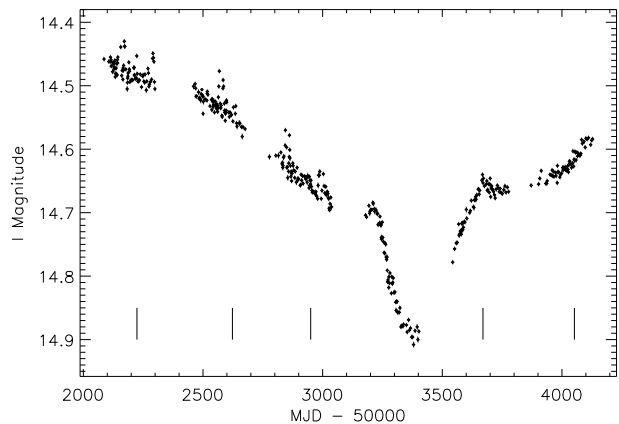}
 \caption{The OGLE III light curve of SXP46.6.  The times when optical spectra 
were taken are marked with solid lines.}
 \label{fig:sxp46_lc}
\end{figure}

\xte\ monitoring is providing a wealth of information on the X-ray pulsars in 
the SMC \citep[see e.g.][]{lay05,gal07}.  However, to investigate the sources 
in detail, in particular at multi-wavelengths, more precise locations must 
be determined, via e.g. \chan\ and \xmm.

In this paper we present observations and analysis of optical and X-ray data
for two Be/X-ray binaries in the SMC, SXP46.6 and SXP6.85, whose positions have
been identified recently using \chan\ and \xmm.

\begin{figure*}
 \includegraphics[width=100mm]{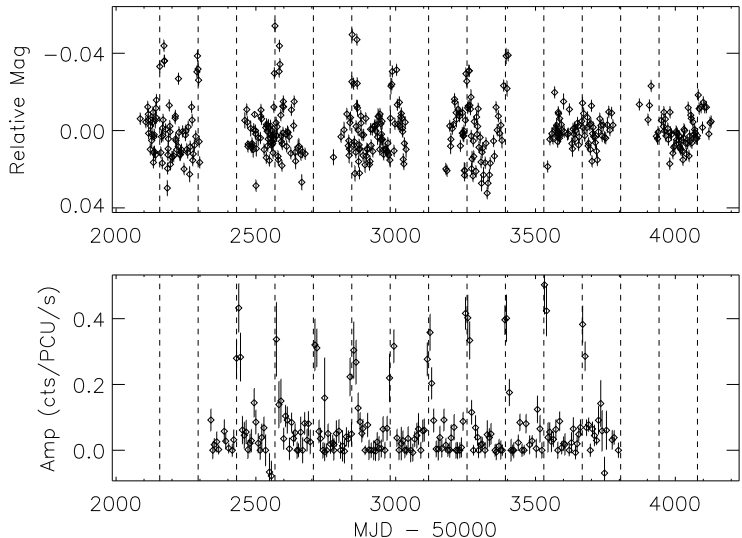}
 \caption{The detrended OGLE III light curve (top) and \xte\ light curve
(bottom) for SXP46.6.  Only X-ray data with identical collimator responses 
(see text) and covering the time of overlap with the optical data have been 
plotted.  The times corresponding to the orbital period determined from the 
\xte\ data, $P=137.74$ d, are marked (dashed lines).}
 \label{fig:sxp46_ox}
\end{figure*}

\section{SXP46.6 = XTE J0053-724 = 1WGA 0053.8-7226}
\label{sect:sxp46}

SXP46.6 was discovered in 1997 during observations of the SMC with \xte\ 
\citep{mar97}.  A position for the source was obtained from follow-up 
{\it ASCA} observations, leading to the object being identified with an 
archival {\it ROSAT} source.  The {\it ASCA} measurements also revealed 
possible pulsations at $\sim 92$ s \citep{cor97}.  Further analysis of the 
{\it ASCA} and \xte\ data established a pulse period for the source of 
46.6 s \citep{cor98}.  Studies of {\it ROSAT} observations confirmed the
transient nature of the source \citep[][and references therein]{buc01}.
Due to the uncertainty on the X-ray position \citet{buc01} proposed two 
optical sources as the counterpart to SXP46.6.  The characteristics of both 
optical candidates were similar and led to the conclusion that the system was 
a Be/X-ray binary.  

\xte\ has continued to monitor SXP46.6 over the past 10 years, the results of
which are presented in \citet{gal07}.  During this time the source has 
exhibited regular outbursts allowing an orbital period of $137.36 \pm 0.35$ d
and an outburst ephemeris of MJD $52293.9 \pm 1.4$ to be derived 
\citep[][based on nine years of data]{gal07}.  Adding the most recent data
we determine a refined period and time of maximum flux of $137.74 \pm 0.42$ d 
and MJD $52430.3 \pm 1.4$, respectively.

\begin{figure}
\begin{center}
 \includegraphics[width=70mm]{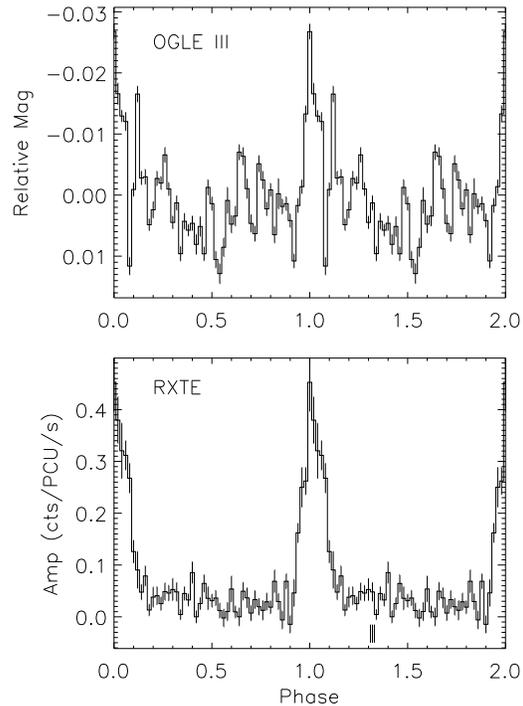}
 \caption{Detrended OGLE III (top) and \xte\ data (bottom) for SXP46.6.  The 
data are folded in 50 phase bins using $P=137.74$ d and $T_{0} =$ MJD 52430.3.
The phases of the three \chan\ observations are marked in the bottom panel 
(solid lines).}
 \label{fig:sxp46_fold}
\end{center}
\end{figure}

We have analysed archival \chan\ ACIS data covering the region of SXP46.6
\citep[see also][]{mcb07}.  The ACIS observations were taken on 2002 July 20, 
2006 April 25 and 2006 April 26 and had exposure times of 7.7 ks, 49.3 ks and 
47.4 ks, respectively.  We processed the data using the \chan\ analysis
package {\sevensize CIAO V}3.4.  The event files were filtered to restrict the 
energy range to 0.5--8.0 keV and we searched for sources using the wavelet 
analysis algorithm.  We detect an X-ray source coincident with one of the 
optical candidates, Star B from \citet{buc01}, in all three \chan\ 
observations.  The X-ray position of the source is RA $=00^{\rm h}$ 
$53^{\rm m}$ $55\fs22$, Dec. $=-72\degr$ $26\arcmin$ $45\farcs7$.  
The measured counts in the three \chan\ observations are 7, 38 and 
28 counts, respectively, corresponding to X-ray luminosities of 
$4.6 \times 10^{33}$, $4.1 \times 10^{33}$ and $3.1 \times 10^{33}$
erg s$^{-1}$, assuming a distance to the SMC of 60 kpc 
\citep[based on the distance modulus,][]{wes97}.  Using the most recent
orbital period and ephemeris from \xte\ we find that all three \chan\ 
observations were taken at outburst phase $\sim0.3$ (see also Figure 
\ref{fig:sxp46_fold}).  The low flux detected with \chan\ can be explained
by the phases at which the measurements occurred.  A search for 
pulsations was not possible due to the lack of source counts.

\subsection{Optical light curve}
\label{sect:sxp46_opt}

\begin{figure}
\begin{center}
 \includegraphics[width=70mm]{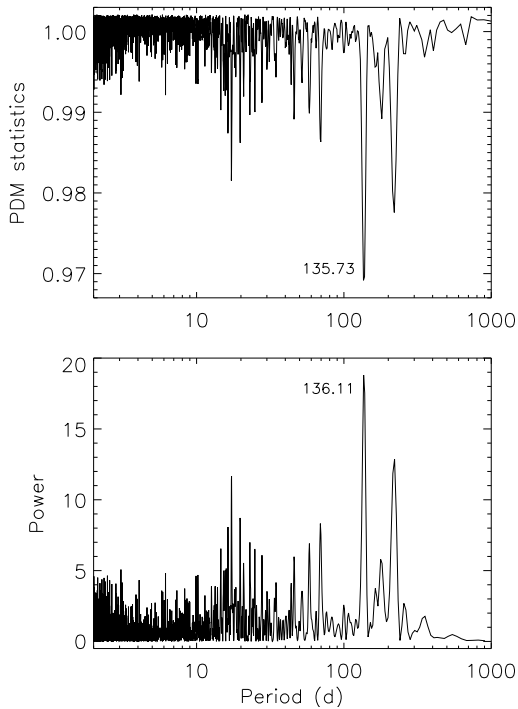}
 \caption{Phase dispersion minimisation (top) and Lomb-Scargle periodograms 
(bottom) for the detrended OGLE III data of SXP46.6.  The strongest dip in the
former corresponds to a period of 135.73 d, with the highest peak at 136.11 d 
in the latter.}
 \label{fig:sxp46_psearch}
\end{center}
\end{figure}

While the source is not in the OGLE II or MACHO catalogues, it is clearly 
detected in the more extensive OGLE III survey.  OGLE III, whose data are not 
yet public, is a continuation of the OGLE project \citep{uda97,szy05}.  We 
show in Figure \ref{fig:sxp46_lc} OGLE III photometry for Star B, from 2001 
June 26 to 2007 January 28.  The optical light curve displays a large 
variation in brightness ($\sim0.5$ mag) over the six years of observations, 
typical of Be stars \citep{men02}.

As the large-scale modulations present in the OGLE III light curve are clearly 
not periodic, before we searched for coherent variations we removed the 
long-term changes.  The data were split into eight sections and each section 
was detrended separately by subtracting a linear fit.  We then searched the 
detrended light curve (see Figure \ref{fig:sxp46_ox}, top panel) for 
periodicities in the range 2--1000 d using Lomb-Scargle (LS) and phase 
dispersion minimization (PDM) periodograms.  

The results of the temporal analysis are shown in Figure 
\ref{fig:sxp46_psearch}.  The dominant peak in the LS periodogram occurs at 
$P=136.11 \pm 2.27$ d, with a corresponding dip in the PDM periodogram at 
$P=135.73 \pm 2.26$ d.  The second largest peak (dip) in the periodograms (at 
$\sim 220$ d) is due to this frequency beating with the one year sampling.
The $\sim 136$ d period from the LS and PDM searches are, within errors, 
consistent with each other and the period found using the most recent 
\xte\ data.  This confirms that Star B is the correct counterpart to SXP46.6.

\begin{figure}
 \includegraphics[width=80mm]{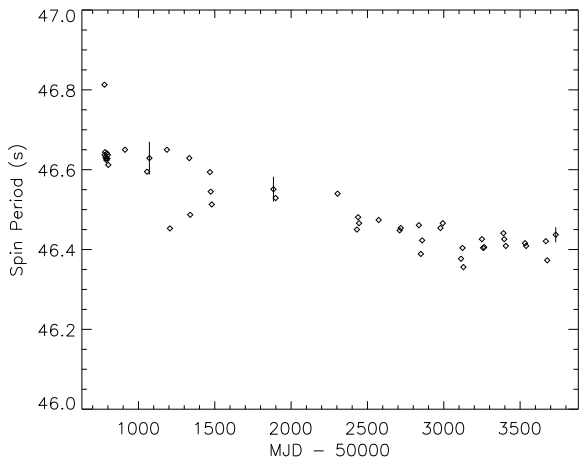}
 \caption{Spin-up of SXP46.6 over the last 10 years, from \xte\ monitoring.}
 \label{fig:sxp46_spin}
\end{figure}

\subsection{X-ray light curve}
\label{sect:sxp46_xr}

Before analysis of the \xte\ light curves can be performed the raw
data are filtered to include only {\it Good Xenon} measurements from the top
anode layer.  Light curves are generated in the 3--10 keV energy band
with 0.01 s time bins and the detected count rate is divided by the number of 
Proportional Counter Units (PCUs) that were active at each time stamp.  Each 
PCU has a collimator, the result of which is that each unit has approximately 
the same field of view.  Due to small variations between the PCUs the source 
count rates must be corrected for the collimator response.  Only observations
with the same pointing position have been used in the following analysis.  For 
more details on the data reduction see \citet{gal07}.

To investigate possible correlations between the optical and X-ray data, we 
show in Figure \ref{fig:sxp46_ox} the detrended OGLE III light curve for 
SXP46.6 (top panel) and the \xte\ light curve (in units of counts PCU$^{-1}$ 
s$^{-1}$, bottom panel).  Only X-ray data with identical collimator responses 
and covering the time of overlap with the optical data have been plotted.  We 
have also plotted, as dashed lines, the times of expected X-ray outburst using 
the orbital period and ephemeris found using the most recent \xte\ data.  It 
is clear from the figure that the optical light curve displays outbursts, with 
amplitudes of $\sim 0.04$ mag, coincident with the X-ray outbursts.

We have folded the detrended OGLE III data and the \xte\ data on $P=137.74$ d 
using $T_{0}=$ MJD 52430.3.  The folded light curves show that 
the modulation is highly non-sinusoidal, with the peak emission in the X-ray 
data coinciding with the peak in the optical data (Figure 
\ref{fig:sxp46_fold}).  It is also apparent that both the X-ray and optical 
outbursts are short-lived, occupying only a small fraction of the orbital
phase.  There is a suggestion that the optical peak is actually split 
in two.  To investigate this we divided the light curve into two parts, 
MJDs 52086--53037 and MJDs 53179--54128, and then folded the two datasets using
the period and ephemeris given above.  We find that the double peak is 
prominent in the first half of the data, with little evidence of the same 
structure in the second half.

\begin{figure}
 \includegraphics[width=84mm]{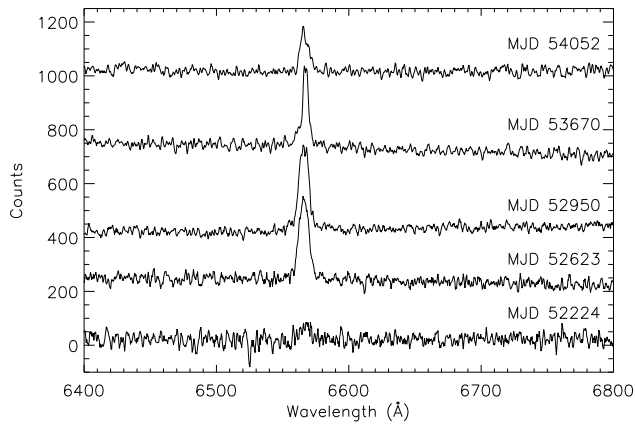}
 \caption{Optical spectra for SXP46.6 taken with the SAAO 1.9 m telescope, 
plotted in order of ascending date.  The spectra have been smoothed with
a boxcar average of three and offset for clarity.  Note the variable H$\alpha$ 
emission.}
 \label{fig:sxp46_spec}
\end{figure}

\subsection{Spin period}

We show in Figure \ref{fig:sxp46_spin} the pulse period history of SXP46.6 
over the past 10 years of \xte\ monitoring.  In general the source has been 
spinning-up over that time frame, with an overall rate of change of spin 
period of $\dot{P}=-1.2 \times 10^{-9}$ ss$^{-1}$.  The transfer of angular 
momentum to the neutron star from orbiting material as it is accreted is 
thought to be responsible for the spin-up \citep{gho79}.  The luminosity 
required to achieve the measured spin-up is $1.2\times 10^{36}$ erg s$^{-1}$.
There is evidence that the rate of spin-up has slowed, with a noticeable 
flattening off in period change, since MJD 53000 (see Figure 
\ref{fig:sxp46_spin}).  This coincides with the lower H$\alpha$ equivalent 
widths measured in the optical spectra (see Section \ref{subsect:spec}) and 
the diminishing size of the optical outbursts (see Figure \ref{fig:sxp46_ox}).

\begin{figure*}
 \includegraphics[width=100mm]{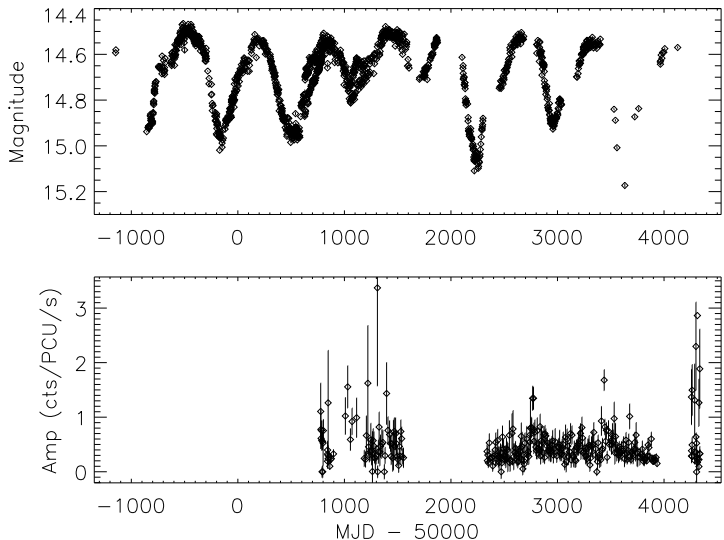}
 \caption{The combined MACHO {\it blue}, OGLE II and OGLE III light curves 
(top) and \xte\ light curve of SXP6.85 (bottom).  The MACHO blue data have 
been shifted by $+24.82$ mag, and the OGLE III data by $-0.3$ mag, both with 
respect to the OGLE II data.  The optical error bars are smaller than the 
plotted symbols.}
 \label{fig:sxp6_lc}
\end{figure*}

\subsection{Optical spectra}
\label{subsect:spec}

Spectra of the optical counterpart of SXP46.6 have been taken on several 
occasions using the 1.9 m telescope at SAAO, South Africa.  A log of the 
observations is given in Table \ref{tab:sxp46_spec}.  The unit spectrograph 
was used with a 1200 line mm$^{-1}$ grating, giving a dispersion of 
$\sim0.4$\AA/pixel.  We also obtained a lower resolution spectrum of the source
using the ESO faint object spectrograph and camera (EFOSC2) mounted on the 3.6 
m telescope at La Silla, Chile.  A 600 line mm$^{-1}$ grating was used giving 
a dispersion of 2\AA/pixel.  The SAAO and ESO data were reduced using standard 
{\scriptsize IRAF} packages.  The SAAO spectra are plotted in Figure 
\ref{fig:sxp46_spec}.  It is clear that the profile of the H$\alpha$ emission 
is changing with time; the equivalent widths are given in Table 
\ref{tab:sxp46_spec}.  The variations in H$\alpha$ are likely due to 
variations in the circumstellar disc of the Be star.

\begin{table}
 \begin{minipage}{80mm}
  \caption{Observation log of the optical spectra for SXP46.6}
  \label{tab:sxp46_spec}
  \begin{tabular}{@{}ccccc}
  \hline
   Date & MJD & Telescope & Exp & H$\alpha$ EW \\
        &     &           & (s) & (\AA)  \\
  \hline
1998-08-19/20 & 51044/5 & SAAO 1.9 m & $2\times600$ & $-15.3\pm0.9$\footnote{\citet{buc01}} \\
2001-11-11 & 52224 & SAAO 1.9 m & 1000 & $-2.6\pm1.0$ \\
2002-12-15 & 52623 & SAAO 1.9 m & 1500 & $-20.2\pm0.5$ \\
2003-11-07 & 52950 & SAAO 1.9 m & 1000 & $-23.2\pm0.6$ \\
2005-10-27 & 53670 & SAAO 1.9 m & 2000 & $-8.3\pm0.4$ \\
2006-11-13 & 54052 & SAAO 1.9 m & 1500 & $-11.2\pm0.5$ \\
2007-09-17 & 54360 & La Silla 3.6 m & 500  & $-12.8\pm0.2$ \\
  \hline								
\end{tabular}							
\end{minipage}							
\end{table}							

\section{SXP6.85 = XTE J0103-728}
\label{sect:sxp6}

X-ray emission, pulsed at 6.8482 s, was first detected from SXP6.85 in 2003
with \xte\ \citep{cor03}.  Continued \xte\ observations spanning two months in 
2003 displayed the transient nature of the source \citep{cor03}.  A further 
five detections have been made during the 10 year monitoring campaign of the 
SMC with \xte\ \citep[see][]{gal07}.  Temporal analysis of the X-ray data by 
\citet{gal07} revealed a period at 112.5 d, but it is not clear if this 
represents a true modulation or if this value is driven by the interval between
two major outbursts.  SXP6.85 was detected in an \xmm\ observation in 2006 
leading to the identification of the source with a $V=14.6$ Be star 
\citep{hab07}.  Independently, \citet{sch07} analysed MACHO and OGLE II data 
for the source finding that the optical brightness varied over a timescale of 
$\sim 658$ d.  They also find a possible period in the OGLE II data of 
$24.82$ d.

\subsection{Optical light curve}
\label{sect:sxp6_opt}

A search of the MACHO \citep{alc99} and OGLE II databases \citep{uda97,szy05} 
provided optical photometry for SXP6.85.  The MACHO project generated 
simultaneous photometry in two passbands, a {\it red} band ($\sim6300-7600$ 
\AA) and a {\it blue} band ($\sim4500-6300$ \AA), both measured in 
instrumental magnitudes.  We have also obtained the OGLE III data for the 
source.  In Figure \ref{fig:sxp6_lc} (top panel) we show the combined MACHO 
{\it blue}, OGLE II and OGLE III light curves of the source, from 1992 August 
18 to 2007 January 28.  In order to display the data on the same scale we had 
to shift the MACHO {\it blue} data by $+24.82$ mag, and the OGLE III data by 
$-0.3$ mag, both with respect to the OGLE II data.

\begin{figure}
 \begin{center}
 \includegraphics[width=70mm]{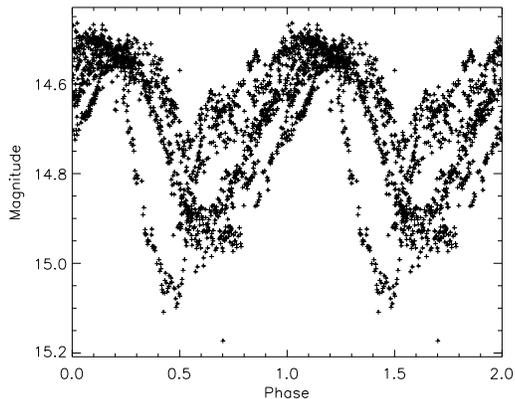}
 \caption{Combined light curve of SXP6.85 folded on $P=620.68$ d.}
 \label{fig:sxp6_lfold}
 \end{center}
\end{figure}

The combined light curve displays large variations in brightness over the 
$\sim$14.5 years of observations.  To determine whether these changes
are periodic we searched the data for coherent modulations in the range 
2--1000 d using LS and PDM periodograms.  Both periodograms show a strong peak
(dip) at $\sim 620 \pm 18$ d, however the folded light curve indicates that 
the variations are quasi-periodic (Figure \ref{fig:sxp6_lfold}).  This is 
similar to the value found by \citet{sch07}.

Using a similar analysis to SXP46.6, we detrended the combined light curve of 
SXP6.85 by splitting it into several sections and then subtracted a linear fit 
from each section.  We then performed a period search with LS and PDM 
periodograms, again over the range 2--1000 d.  Both periodograms show several 
large peaks (dips), with the highest peak in the LS periodogram occurring at
$114.07\pm 0.62$ d, and the corresponding PDM dip at $114.01\pm0.62$ d.  
There is a smaller peak (dip) at 25.86 d which is close to the value reported 
by \citet{sch07}.  However, due to the large aperiodic changes in the raw data
it is likely that these peaks (dips) are spurious and do not represent true 
periodic behaviour.  We find that the amplitudes of the variations in the 
detrended light curves folded on these values are $<0.01$ mag.

\begin{figure}
 \includegraphics[width=84mm]{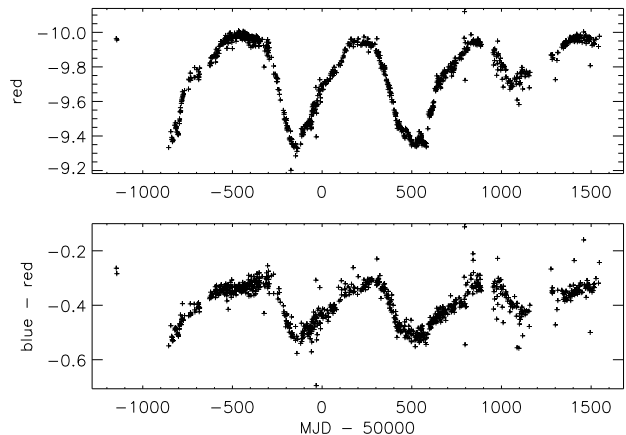}
 \caption{MACHO {\it red} light curve (top) and MACHO $blue - red$ colour 
variations over time (bottom) for SXP6.85.  Note that the source gets redder 
as it gets brighter.}
 \label{fig:sxp6_col}
\end{figure}

\subsection{X-ray light curve}
\label{sect:sxp6_xr}

The \xte\ light curve of SXP6.85 was generated in the same way as for SXP46.6
(see Section \ref{sect:sxp46_xr}).  We show in Figure \ref{fig:sxp6_lc} 
(bottom panel) the \xte\ light curve of SXP6.85.  There does not appear to be 
any obvious correlation between the X-ray outbursts and the raw optical data 
(Figure \ref{fig:sxp6_lc}, top panel) or with the detrended optical data (not 
shown).

\subsection{Colour variations}
\label{sect:sxp6_col}

We show in Figure \ref{fig:sxp6_col} how the colour of SXP6.85 varies over 
time compared to the MACHO red data.  It can be seen that as the source 
brightens it gets redder.  This indicates that the variations in the light 
curve are likely to be caused by changes in the structure of the circumstellar 
disc around the Be star.  This is due to the disc being redder in $B-V$ (i.e. 
cooler) than the Be star \citep{jan87}.  Thus, the formation of the 
circumstellar disc will increase the optical brightness of the system by the 
addition of red light, or it will make the system appear fainter by masking 
the Be star, behaviour that is dependant on the inclination of the system.
It is interesting to note that the shape of the colour changes do not exactly 
mirror the variations in the red light curve.

\section{Discussion}
\label{sect:disc}

\subsection{SXP46.6}

The optical light curve of SXP46.6 displays long-term variations that are
most likely due to the formation and depletion of the Be star's circumstellar 
disc.  It is reasonable to assume, if the equatorial plane of the Be star
coincides with the orbital plane of the compact object, that as the neutron 
star orbits close to this disc at periastron X-ray outbursts should be 
observed.  Our results show that not only do we detect X-ray outbursts, but 
optical outbursts are also seen.  SXP46.6 is therefore one of only a handful 
of systems that displays optical bursts at the binary period 
\citep[see e.g.][]{alc01,sch03,cow03,coe04,edg05}.  We find that the amplitude 
of the optical outbursts ($\sim 0.04$ mag) is similar to those of other 
sources \citep[see e.g.][]{sch03,cow03}.

In Be/X-ray transients in general, the depletion of the circumstellar disc
leads to a decrease in the optical brightness of the system.  For SXP46.6, 
even when the optical light curve goes into a sharp decline there must be a 
residual disc present as outbursts, both at optical and X-ray wavelengths, are 
seen.  In contrast, as the optical brightness recovers, which would normally 
be attributed to the disc reforming, an X-ray outburst is evident but there 
does not appear to be a corresponding burst in the optical.  The reason for 
this is unclear, but could possibly be related to the amount of material in 
the circumstellar disc available for accretion and the quantity that is 
transferred to the neutron star.  If less matter is accreted the optical 
outbursts could have a lower amplitude and may not be detectable above 
the intrinsic variability in the optical light curve.  Another possibility for the 
lack of an optical outburst at this time is that there remains a wind that is
sufficient to generate X-ray outbursts only.

Using the H$\alpha$ equivalent width as an indicator for the circumstellar 
disc size \citep{dac86}, we find that for the first three years of OGLE III 
monitoring the disc size is increasing, anti-correlated with the optical 
brightness.  Unfortunately we do not have spectra during the optical
minimum.  After the optical minimum the equivalent width of the H$\alpha$ 
emission line seems to reflect the recovery of the circumstellar disc, with the 
width of the line increasing as the optical brightness increases.  The 
anti-correlated behaviour initially may be due to the disc covering part of the 
Be star leading to a reduction in brightness.  This has been seen in another  
source, A 0538-66 \citep{alc01}.  It is unfortunate that we do not have 
multi-band data that would allow us to investigate the colours of the source 
at these times.

The shape of the optical and X-ray outbursts for SXP46.6 are both asymmetric, 
with a faster rise and slower decline, reminiscent of other Be/X-ray binaries 
\citep[see e.g.][]{alc01,coe04}.  We propose that the optical bursts could 
be due to the circumstellar disc being perturbed at each passage 
causing the disc to grow slightly, leading to an enhancement in the optical 
brightness.  We can calculate a rough estimate for this increase in 
brightness.  If we assume that the 0.5 magnitude global change in the light 
curve represents the difference between no disc and a full disc, then this 
corresponds to an increase in light of 58\%.  The optical outbursts we detect 
have an amplitude of $\sim 0.04$ mag, which equals a total increase from no 
disc of 64\%.  Assuming that the increase in brightness is equivalent to the 
disc size, then during outburst the disc grows by $\sim 10$\%.  We note 
that the increase in optical brightness at outburst may not be due to 
geometric effects alone.  We find that the luminosity of the optical 
outbursts is $\sim3\times10^{34}$ erg s$^{-1}$ and the luminosity of the 
X-ray outbursts is $\sim2\times10^{36}$ erg s$^{-1}$.  This implies that 
reprocessing could be contributing to the change in the optical light.

\citet{stel86} distinguished between the various kinds of X-ray behaviour 
observed from Be/X-ray binaries, placing them into three groups: (i) 
persistent, low-level (quiescent) emission, (ii) relatively bright, 
short-lived outbursts recurring on the orbital period at the time of 
periastron passage, known as Type I, and (iii) large, long-lived (weeks to 
months) outbursts that occur irregularly, known as Type II.  We are most likely
detecting quiescent emission from SXP46.6 in the \chan\ observations.
\citet{oka01} have shown that the class of X-ray outbursts detected from a 
Be/X-ray transient, Type I and/or Type II, is dependent on the orbital 
eccentricity of the system, and the viscosity of the disc.  The disc in 
binaries with low eccentricity is truncated at the 3:1 resonance radius.  This 
means that too little material is captured (if any) by the compact object as 
it orbits the Be star as the disc is too far away.  Therefore, regular Type I 
outbursts cannot be produced by these systems, but they may show infrequent 
Type II outbursts.  In contrast, periodic Type I outbursts are generated in 
high eccentricity binaries as the disc is truncated at a larger resonance 
radius, which permits material to be accreted by the neutron star every 
orbit.  The detection of Type I X-ray outbursts from SXP46.6, and the 
corresponding optical outbursts, and its similarity to other sources that have 
shown optical outbursts on the orbital period, implies that SXP46.6 has a high 
orbital eccentricity.

\subsection{SXP6.85}

The long-term optical light curve of SXP6.85 is similar to that of SXP46.6 in 
the fact that the large changes observed are not periodic, leading to the 
conclusion that they are related to the Be phenomenon.  The variations in the 
source's colour support this.  Even after detrending, we find little evidence 
for any coherent modulations in the optical light curve of SXP6.85.

Analysis of SXP6.85's X-ray light curve does not firmly demonstrate the 
presence of periodic behaviour either \citep{gal07}.  More data are needed to
confirm the tentative orbital period proposed.  There is a lack of correlation 
between the optical and X-ray data.  By comparing the two datasets we find 
that one X-ray outburst occurs just prior to optical maximum, while another 
two seem to occur as the optical source is fading.  This implies that the
neutron star is not accreting directly from the stellar wind, and could
indicate that the X-ray emission is triggered by discrete episodes of 
disc material being lost radially \citep[see e.g.][]{cla99}.

The characteristics of SXP6.85 are very similar to those of EXO 0531-66 and 
H 0544-665 \citep{mcg02}.  All three sources are highly variable in the 
optical over long timescales but do not show variations that are orbital 
in origin, they become redder as they brighten in the optical and no strong
correlation is found between the X-ray and optical measurements.  Due to the
similarities of these sources, and with comparison to SXP46.6, we therefore
propose that SXP6.85 is a low eccentricity system.

\section{Summary}

We find that the long-term optical light curve of SXP46.6 displays regular 
outbursts that are coincident with the X-ray outbursts, both modulated on a 
period of $\sim 137$ d.  The behaviour of SXP46.6 indicates that it has a 
high orbital eccentricity.  Our results from optical spectra of SXP46.6
imply that the Be star's circumstellar disc varies in size over the years.
We do not find any periodic modulations in the optical light curve of SXP6.85,
but our analysis does show that as the source brightens it gets redder.  It is
therefore likely that the changes in the light curve and the colour variations
are due to the formation and depletion of the circumstellar disc.  The lack of 
correlated X-ray and optical activity for SXP6.85 suggests that it is a low
eccentricity system.

\section*{Acknowledgments}

Support for OGLE was provided by the Polish MNiSW grant N20303032/4275.
This paper utilizes public domain data originally obtained by the MACHO 
Project, whose work was performed under the joint auspices of the U.S. 
Department of Energy, National Nuclear Security Administration by the 
University of California, Lawrence Livermore National Laboratory under 
contract No. W-7405-Eng-48, the National Science Foundation through the Center 
for Particle Astrophysics of the University of California under cooperative 
agreement AST-8809616, and the Mount Stromlo and Siding Spring Observatory, 
part of the Australian National University.

\label{lastpage}

\end{document}